\documentclass[aps,pre,twocolumn,groupedaddress]{revtex4}

\def\sR{\hbox{I\kern-.1667em\hbox{R}}}

\def\vn{{\bf n}}

\def\ep{\epsilon}

\def\beq{\begin{equation}}
\def\eeq{\end{equation}}
\def\bea{\begin{eqnarray}}
\def\eea{\end{eqnarray}}

\def\half{{\textstyle{1\over 2}}}

\def\etal{{\it et al.}}

\begin{document}


\title{Bose-Einstein Condensates as a Probe for Lorentz Violation}


\author{Don Colladay}
\email[]{colladay@ncf.edu}
\affiliation{New College of Florida}

\author{Patrick McDonald}
\email[]{mcdonald@ncf.edu}
\affiliation{New College of Florida}


\date{\today}

\begin{abstract}
The effects of small Lorentz-violating terms on Bose-Einstein condensates
are analyzed.  We find that there are changes to the phase and shape of the 
ground-state wave function that vary with the orientation of the trap.  In 
addition, spin-couplings can act as a source for spontaneous symmetry breaking
in ferromagnetic condensates making them sensitive probes for fundamental 
symmetry violation.

\end{abstract}


\maketitle

\section{Introduction}

Recently there has been much interest in searching for miniscule violations
of Lorentz symmetry in nature.  These effects may arise from more fundamental theories
that underly the standard model.
Bose-Einstein condensates \cite{dalfovo} have provided a rich testing ground for large, coherent quantum
mechanical systems.
By now, condensates have been successfully produced using a number of
atomic species in different spin states using a variety of trapping techniques.
It is the goal of this paper to analyze the general effects of small Lorentz-breaking
terms on these condensates.

Tests of Lorentz symmetry have been performed in a wide variety of 
physical systems \cite{cpt98}.  
For example, various experiments utilizing mesons \cite{mesons1,mesons2,mesons3}, 
baryons \cite{baryons1,baryons2,baryons3}, electrons \cite{electrons1,electrons2,electrons3},
photons \cite{ph1,ph2,ph3,ph4,ph5}, and muons \cite{muons} have reached a precision that probes
Lorentz-violating parameters at Planck-suppressed scales.
Recent analysis has also been extended to the neutrino sector \cite{neutrinos1,neutrinos2},
instantons \cite{instantons}, supersymmetric models \cite{berger},
baryogenesis \cite{bckp},
and the gravitational sector \cite{kgrav}.

A framework for including Lorentz-breaking effects into low-energy field
theory is provided by the Standard Model Extension (SME).
The SME uses the concept of spontaneous symmetry breaking to generate
couplings between standard model fields and vacuum expectation values
of tensor fields that parameterize the symmetry violations \cite{kps}.
In this paper, a more tractable version of the SME is used that includes 
various restrictions on the possible couplings that preserve gauge invariance 
and power counting renormalizability \cite{ck}.
This restricted theory has been shown to preserve microcausality in concordant frames
where the Lorentz-violating terms are reasonably small \cite{ralph}.

Previous related work involves an analysis of statistical mechanics in the 
presence of Lorentz violation \cite{statmech} and provides the formal thermodynamic
techniques used in the current paper.
One important modification is the explicit inclusion of the confining potential,
necessary for describing a condensate wave function.

The paper is organized as follows:
In section II, the simplified case of spin-0 bosons in a condensate is 
studied.  This gives the basic effects on the spin-independent part of the
wave function.
Section III generalizes the system to spin-polarized hydrogen
(in the absence of interactions), a relatively simple
physical boson used in experiments.
Section IV includes the effects of interactions and the more complex atoms used in 
the majority of experiments.
Section V generalizes to optically-trapped condensates with multiple possible
spin components.

\section{Noninteracting Spin-0 Bosons}

As a first step, we consider the case of noninteracting spin-0 bosons in
a harmonic trap
(interactions and spin effects are discussed in later sections).
A free spin-0 boson gas in the presence of Lorentz violation 
may be modelled using the effective
hamiltonian \cite{statmech}
\begin{eqnarray}
H & = & \frac{p^2}{2m} + A +  C_{j}\frac{p_j}{m} +
F_{jk}\frac{p_jp_k}{2m} + V_{\rm trap}
\quad ,
\label{H3}
\end{eqnarray}
where $V_{\rm trap}$ is the trapping potential \cite{fn2}.
The parameters $A$, $C$, and $F$ are effective Lorentz violation
parameters for the boson.  They are expressible in terms of 
fundamental violation parameters of the electrons, protons, and
neutrons, if the explicit wave function
for the boson is known.

In a previous paper \cite{statmech}, a calculation of the properties 
of a low-temperature
Bose gas was performed and the trapping potential
was taken as a standard particle in a box.  
When a condensate is present, detailed knowledge of the trapping potential is 
required to calculated the 
ground state wave-function.
For a large class of magnetic and optical traps, the trapping potential takes the
form of a harmonic oscillator
\beq
V_{\rm trap} = {1 \over 2} m \sum_i (\omega_i x_i)^2
\quad .
\eeq
The ground state wave function plays a special role, 
therefore it is useful to first study some of it's properties.
Without Lorentz violation, the unperturbed ground state takes the standard form
\beq
\psi^0_{\rm GS} = \left( {m \omega_{\rm ho} \over \pi \hbar}\right)^{3/4} 
\exp{(-{m \over 2 \hbar}\sum_i \omega_i x_i^2)}
\quad ,
\eeq
where $\omega_{\rm ho} = (\omega_1 \omega_2 \omega_3)^{1/3}$.
The unperturbed energy for the ground state is
\beq
E^0_{\rm GS} = {1 \over 2} \hbar (\omega_1 + \omega_2 + \omega_3)
\quad .
\eeq
The first order correction to the energy can easily be found using standard
first-order perturbation theory as
\beq
E^1_{\rm GS} = \langle \psi^0_{\rm GS}|H^\prime |\psi^0_{\rm GS}\rangle = 
A + {1 \over 4} \hbar \sum_i F_{ii}\omega_i
\quad .
\eeq
Similarly, the first-order correction to the ground-state wave function
is found using
\beq
\psi^1_{\rm GS} = \sum_{\vec n \ne 0} {\langle \psi_{\vec n}^0|H^\prime|\psi^0_{\rm GS} \rangle 
\over E^0_{\rm GS} - E^0_{\vec n}}\psi^0_{\vec n}
\quad ,
\eeq
where $\psi^0_{\vec n}$ are the standard unperturbed states that can be written in
terms of appropriate Hermite polynomials and exponentials.
Only a few of the matrix elements in the sum are nonzero.  
The resulting corrections are calculated first for the $C$-terms as
\bea
\psi^{(C)1}_{\rm GS} & = & -i ({1 \over 2 m \hbar})^{1/2}[C_1 \omega_1^{-1/2} \psi^0_{100} + 
C_2 \omega_2^{-1/2} \psi^0_{010} \nonumber \\
& & 
+ C_3 \omega_3^{-1/2} \psi^0_{001}] \nonumber \\
& = & - {i \over \hbar} \vec C \cdot \vec x \psi^0_{\rm GS}
\quad .
\eea
In fact, the exact solution for this case is given by
\beq
\psi^{(C)}_{\rm GS} = \exp{(-i {\vec C \cdot \vec x \over \hbar})} \psi^0_{\rm GS}
\quad ,
\eeq
indicating that the sole effect is to introduce a position-dependent
phase shift into the ground state.
Such a term could contribute to the pattern observed in
condensate interference experiments \cite{interf}.

The corrections due to the $F$-terms may be handled using the same procedure.
The resulting first-order correction for the ground state is
\bea
\psi^{(F)1}_{\rm GS} & = & {1 \over 2}\left[ {1 \over 2 \sqrt{2}}(F_{11}\psi^0_{200} +
F_{22}\psi^0_{020} + F_{33}\psi^0_{002}) \right. \nonumber \\
& & \left. + {(\omega_1 \omega_2)^{1/2} \over \omega_1 + \omega_2}F_{12}\psi^0_{110}
+ {(\omega_2 \omega_3)^{1/2} \over \omega_2 + \omega_3}F_{23}\psi^0_{011} \right. \nonumber \\
& & \left. + {(\omega_1 \omega_3)^{1/2} \over \omega_1 + \omega_3}F_{13}\psi^0_{101}
\right] 
~ .
\eea
Substitution of the unperturbed states yields the explicit form
\beq
\psi^{(F)1}_{\rm GS} = {m \over 2 \hbar} \left[
\sum_{i, j} {\omega_i \omega_j \over \omega_i + \omega_j} F_{ij} x_i x_j
\right] \psi^0_{\rm GS} ~ .
\eeq
This shows that the condensate shape takes the form of a perturbed ellipsoid.

The Fourier transform gives the momentum-space wave function as
\bea
\phi_{\rm GS}(\vec p) & & = \left( {1 \over \pi \hbar m \omega_{\rm ho}}\right)^{3/4}
(1 + {1 \over 8} Tr(F)) \otimes
\nonumber \\
& & \exp{\left(-{1 \over 2 m \hbar} \left[ 
\sum_i {p_i^2 \over \omega_i} + \sum_{i, j} F_{ij}
{p_i p_j \over \omega_i + \omega_j}\right] \right)} ~ .
\eea
This formula provides the momentum distribution of the particles in the 
condensate.  
If the trapping potential is suddenly turned off, the velocity distribution
can be measured and compared with the above formula.
Unfortunately, current shape sensitivity is only at the 1 \% level 
\cite{hau} and is unlikely to yield interesting bounds on Lorentz violation
parameters in the near future.
We now turn to the case of finite temperature and analyze the particle
distribution.

Employing the notation of \cite{statmech} 
the associated grand partition function is
\begin{eqnarray}
\ln Z_G & = & -\sum_\vn \ln(1-e^{-\alpha}e^{-\beta
E_{\vn}}) \nonumber \\
& & - \ln(1-e^{-\alpha - \beta E_{\rm GS}}) \quad ,
\end{eqnarray}
where $\alpha = - \beta \mu$ is defined in terms of the chemical potential,
and the ground state has been separated out to allow for
Bose-Einstein condensation at low temperatures. 
In order for a large number of atoms to condense into the
ground state, the chemical potential must be very close to 
the ground state energy.
It is therefore convenient to define $\mu = E_{\rm GS} - \epsilon$, where
$\epsilon$ is a small parameter.

Approximating the sum as an integral and taking the limit as 
$\epsilon$ gets small gives the result
\begin{equation}
\ln Z_G =  (1 - \half Tr(F)) \left( {k T \over \hbar \omega_{\rm ho}}\right)^3
I_4(\epsilon) +
\ln(1-e^{-\beta \epsilon})  
~ ,
\end{equation}
where 
\beq
I_\nu(\epsilon) \equiv {1 \over \Gamma(\nu)} 
\int_0^\infty dx {x^{\nu -1} \over e^{- \beta \epsilon + x} - 1} ~,
\eeq
is an integral that reduces to the Riemann Zeta function $I_\nu(\epsilon) \rightarrow \zeta(\nu)$
in the limit $\epsilon \rightarrow 0$.
Note that the thermally distributed particles are only affected by
the rotationally invariant parameter $Tr(F)$ as is expected from
equipartition of energy.

The expected number of particles in excited states can be found by
differentiation of the first term in the partition function with respect 
to the parameter $\alpha = -\beta \mu$.
The result is 
\beq
\langle N \rangle - \langle N_0 \rangle = (1 - \half Tr(F))
\left( {k T \over \hbar \omega_{\rm ho}}\right)^3 I_3(\epsilon)~.
\eeq
The number of particles in the ground state can be found by differentiation
of the second term in the partition function.  The result is
\beq
\langle N_0 \rangle = {1 \over e^{\beta \ep} - 1} ~.
\eeq
When the condensate is present, $\langle N_0 \rangle \simeq kT / \epsilon$ must be large.

The critical temperature for condensation can be found by setting
the number of particles in excited states equal to the total number
of particles in the limit $\epsilon \rightarrow 0$ yielding
\beq
k T_c = \hbar \omega_{\rm ho} (1 + {1 \over 6} Tr (F))N^{1/3} \zeta^{-1/3}(3)
~.
\eeq
Combining the above results gives the relation
\beq
{N_0 \over N} = 1 - (T/T_c)^3
~,
\eeq
showing that the fraction of atoms in the condensate (expressed in
terms of $T_c$) is independent
of the Lorentz-violating parameters.
Note that systems with $T \ll T_c$ are in an almost pure condensate state.
These are the systems focused on in the remainder of the paper.

This completes the description of spin-0 condensates in the presence of 
Lorentz violation.
Most actual experiments involve atoms with nontrivial total spin,
significant interactions, or both.
As a next step, the previous case is generalized to spin-polarized hydrogen.

\section{Spin-Polarized Hydrogen}
Hydrogen provides a theoretically simple example of a physical Bose-Einstein condensate.
Interactions are still neglected; they are in fact important and will be discussed 
in the next section.
The Lorentz-violating terms in the hamiltonian for the system can be taken as a
simple sum of the electron and proton terms.
The momentum terms for the electron and proton may be written in terms of the
total momentum $P$ and the relative momentum $p_r$ in the standard way
\beq
\vec p_e = {\mu_r \over m_p}\vec P + \vec p_r 
~,~
\vec p_p = {\mu_r \over m_e} \vec P - \vec p_r
~,
\eeq
where $\mu_r$ is the reduced mass.
The part of the hamiltonian that is relevant for condensate corrections is
\begin{eqnarray}
H_{LV} & \supset & A^{(e)} + A^{(p)} + B^{(e)}_j \sigma^j_{(e)} + B^{(p)}_j \sigma^j_{(p)} \nonumber \\
& & + (C^{(e)}_{j} + C^{(p)}_j +
D^{(e)}_{jk} \sigma^k_{(e)} + D^{(p)}_{jk} \sigma^k_{(p)})\frac{P_j}{M} \nonumber \\ 
 &  &  + (F^{(e)}_{jk} + F^{(p)}_{jk} + G_{jkl}^{(e)} \sigma^l_{(e)} +
 G_{jkl}^{(p)} \sigma^l_{(p)}) \frac{P_jP_k}{2M} 
\quad , 
\label{hydrogen}
\end{eqnarray}
consisting of the terms that couple to the total atomic momentum.
In this expression, $M$ is the mass of the atom and the superscripts $(e)$ and $(p)$ denote the
parameters for the electron and proton.
The SME parameters for Lorentz violation \cite{klane} have been collected as
\begin{equation}
A = (a_0-mc_{00} - me_0) \quad ,
\end{equation}
\begin{equation}
 B_j = (-b_j + md_{j0} -
\half m \ep_{jkl}g_{kl0} + \half \ep_{jkl}H_{kl})
\quad ,
\end{equation}
\begin{equation}
C_j = [a_j - m(c_{0j} + c_{j0}) - m e_j]
\quad ,
\end{equation}
\begin{eqnarray}
D_{jk} =& & \left[-b_{0}\delta_{jk} + m(d_{kj}+ d_{00}\delta_{jk})
\right. \nonumber \\
& & \left. + m\ep_{klm}(\half g_{mlj} + g_{m00}\delta_{jl}) +
\ep_{jkl}H_{l0}\right]  , 
\end{eqnarray}
\begin{equation}
F_{jk} = - 2 \left[(c_{jk} +
\half c_{00}\delta_{jk})\right]
\quad ,
\end{equation}
\begin{eqnarray}
G_{jkl} & = & 2 \left\{\left[(d_{0j} + d_{j0}) \right. \right.\nonumber \\
& - & \left. \left.\half (b_j/m + d_{j0} +
\half \ep_{jmn}(g_{mn0} + 
H_{mn}/m))\right]\delta_{kl}\right. \nonumber \\ 
 & + & \left. \half (b_l/m
+ \half \ep_{lmn}g_{mn0})\delta_{jk} \right. \nonumber \\
& - & \left. \ep_{jlm}(g_{m0k} +
g_{mk0})\right\} \quad .  
\end{eqnarray}
Note that if the trap selects out the singlet configuration, the system
would be equivalent to the spin-0 case discussed in the previous section.

In order to trap hydrogen magnetically, it must be in the triplet spin configuration.
For example, suppose it is the $|F,m_F \rangle = |1,1\rangle$ state that is trapped.
The spin-couplings contribute to the perturbed energies and ground state and the
calculation is the same as before with the replacement $A \rightarrow A + B_3$
and $F_{ij} \rightarrow F_{ij} + G_{ij3}$.
With this replacement, the correction to the ground state energy is
\bea
E^1_{\rm GS} & = & \langle \psi^0_{\rm GS}|H^\prime |\psi^0_{\rm GS}\rangle 
\\ \nonumber
& = & 
\sum_{e,p}\left[A + B_3 + {1 \over 4} \hbar \sum_i (F_{ii}+G_{ii3})\omega_i
\right]
\quad ,
\eea
where the first sum indicates a sum over proton and electron couplings.
The other case: $|F,m_F \rangle = |1,-1\rangle$ can be easily found by flipping
the signs of the spin-couplings $B$ and $G$.
This example demonstrates explicitly how the corrections can be expressed in 
terms of fundamental SME parameters.  This calculation neglects
interactions that play a significant role in this system.  These are discussed in the
next section.

\section{More Complex Atoms and Interactions}
Most traps use more complicated atoms, such as $^7$Li,
$^{23}$Na, and $^{87}$Rb.
The hamiltonian given in Eq.~(\ref{hydrogen}) can be extended by formally
performing a sum over all constituent particles \cite{klane2}.  
In practice, the resulting hamiltonian is unwieldy and certain approximations
must be made to obtain tractable results.
These three commonly utilized atoms have a nuclear spin of 3/2 with a single 
valence electron.
It is therefore possible to magnetically trap them in a spin-1 or spin-2 state.
The detailed contribution of the various particle types to the ground-state 
corrections will depend on the specific nuclear model used.
One approach is to adopt a Schmidt model in which all of the nuclear spin
is attributed to a single unpaired nucleon.  
For the atoms listed above, the unpaired nucleon is a proton, indicating that 
these atoms are particularly sensitive to proton violation parameters
(this will be discussed further in the section regarding optical traps where 
spin-couplings are important). 
In addition to more complicated hamiltonians, these atoms have significant
interactions that we will now discuss.

In the conventional case with no Lorentz violation, 
at low energies and densities relevant to the condensate, the two-body interactions may be
incorporated using a single parameter $a$, called the scattering length.
The arguments leading to the above conclusion do not depend on the
specific details of the potential between the atoms.
This can be seen by looking at the Born approximation for the scattering amplitude
at low energies (called the scattering length)
\beq
a \simeq -{m \over 4 \pi \hbar^2} \int V(\vec r)d^3 \vec r
~ .
\eeq
As a result, any Lorentz-violating effects in the interaction potential will be
absorbed into the definition of the scattering length.

The second-quantized hamiltonian may be written as
\bea
\hat H & = & \int d^3 \vec r \psi^\dagger(\vec r)\left[ -{\hbar^2 \over 2 m} \nabla^2 
+ H_{\rm LV} + V_{\rm trap} (\vec r) \right] \psi(\vec r)
\nonumber \\
& & + \half \int \int d^3 \vec r d^3 \vec r^\prime \psi^\dagger(\vec r) 
\psi^\dagger(\vec r^\prime)V(\vec r - \vec r^\prime) \psi(\vec r) \psi(\vec r^\prime) 
~,
\nonumber \\
\eea
where $H_{\rm LV}$ is the Lorentz-violating piece of the hamiltonian
and $V(\vec r)$ is the interatomic potential.
This potential may be replaced by the effective interaction
\beq
V(\vec r) = {4 \pi \hbar^2 a \over m} \delta^3(\vec r)
~,
\eeq
because it produces the same scattering behavior as the full potential at low energies
and densities.
The bosonic field operators may be expanded about the condensate wave
function $\Phi$ as $\Psi(\vec r,t) = \Phi(\vec r,t) + \Psi^\prime(\vec r, t)$,
yielding the modified Gross-Pitaevskii equation for the condensate
\beq
i \hbar {\partial \over \partial t} \Phi = \left[ - {\hbar^2 \over 2 m} \nabla^2 
+ H_{\rm LV} + V_{\rm trap} + {4 \pi \hbar^2 a \over m} |\Phi|^2 
\right]\Phi
~.
\eeq
The time dependent piece in the context of mean field theory is given by 
$\Phi(\vec r, t) = \phi(\vec r) \exp{(-i \mu t / \hbar)}$
in terms of the chemical potential $\mu$ as a result of Anderson's equations\cite{and}.
This equation is nonlinear and generally must be solved numerically,
however, the Thomas-Fermi limit is relevant for most experiments
for which the interaction energy is much larger than the kinetic energy
over the bulk of the condensate.
In this limit, the kinetic terms are neglected and the only unsuppressed contribution from
the Lorentz-violating terms comes from the momentum independent spin couplings.
In the case of a strong external magnetic trapping field, the condensate will consist of
a single spin-component, and the density is given by
\beq
n(\vec r) = \phi^2(\vec r) = {m \over 4 \pi \hbar^2 a} (\mu - E_{\rm LV} - V_{\rm trap}(\vec r))
~,
\eeq
where $E_{\rm LV} = \langle \phi |H^{p-indep}_{\rm LV}| \phi \rangle$ is the expectation value of
the momentum-independent terms in the Lorentz-violating hamiltonian.
The field $\phi$ is normalized to the total number of particles in the
condensate such that $\int d^3 \vec r \phi^2 = N_0$,
implying that
\beq
\mu - E_{\rm LV} = {\hbar \omega_{\rm ho} \over 2} \left(
{15 N_0 a \over a_{\rm ho}}\right)^{2/5}
~,
\eeq
where $a_{\rm ho} = (\hbar/ m \omega_{\rm ho})^{1/2}$ corresponds to the average
width of the free-particle condensate solution.
This means that the Lorentz-violating terms may be effectively absorbed into the
chemical potential and therefore do not affect the bulk properties of the 
condensate.

\section{Optical Traps}

More interesting are the nontrivial spin-states that are found in optical traps where a
superposition of various spin projections in the condensate are possible.
The trapping potential is produced using the electric field of an optical beam.
Depending on the scattering lengths for the different spin channels, the condensate
may be ferromagnetic or polar.
In these traps, the Lorentz violation terms coupling to spin can mimic 
external magnetic fields. 
The common case of $f=1$ bosons is considered here.

The Gross-Pitaevskii equation is modified to include spin-dependent
scattering lengths (in the Thomas-Fermi limit) by writing the energy functional 
\beq
K = \int d^3 \vec r n[V_{\rm trap} + {c_0 n \over 2} + {c_2 n \over 2}
\langle \vec F \rangle^2 + E_{\rm ze}]
~,
\eeq
where $c_0$ and $c_2$ are appropriate linear combinations of scattering lengths
for the total spin-0 and total spin-2 scattering states\cite{ho}, and
$E_{\rm ze}$ is the Zeeman energy contributed by any external magnetic
field that may be present.
This expression assumes that angular momentum can be exchanged 
between the condensate and the environment.  A Lagrange multiplier can
be included to incorporate total spin conservation when necessary.
For our purposes, it suffices to set the external magnetic fields to zero, eliminating
the Zeeman contributions to the energy.
In addition, we have neglected any magnetostatic interactions between the boson magnetic
moments.

The Lorentz-violating terms may then be included using the Schmidt model for
the nuclei.  
Assumming that the momentum-dependent terms are suppressed, the only relevant terms are
$\vec B^{(e)}\cdot \vec \sigma_{(e)}$ and $\vec B^{(p)}\cdot \vec \sigma_{(p)}$.
Calculating the matrix elements of these operators in the basis of states
used in the above energy functional yields
the effective Lorentz-violating correction to the hamiltonian for
a single atom
\beq
H_{\rm LV} = - {1 \over 2}\vec B^{(e)}\cdot \vec F + {5 \over 6} \vec B^{(p)}\cdot \vec F
\quad .
\eeq
The proton orbital angular momentum has been set to $l=1$ for definiteness.
The other choice of $l=2$ simply alters the coefficient on the proton term to -1/2.
The spin-dependent part of the energy functional to be minimized is therefore
\beq
K_s = {c_2 n \over 2} \langle \vec F \rangle^2 
+ \left({5 \over 6} \vec B^{(p)} - {1 \over 2}\vec B^{(e)}\right) \cdot \langle
\vec F \rangle ~.
\eeq
The type of condensate formed depends on the sign of $c_2$.
If $c_2>0$, $\langle \vec F \rangle = 0$ minimizes the energy and the state is
called polar.  
If $c_2<0$, $\langle \vec F \rangle \ne 0$ such that $\langle \vec F \rangle^2 = 1$
minimizes the energy.  In this case, the state is called ferromagnetic due to the
collective polarization of the system.
The above expression demonstrates that the effect of the Lorentz-violating terms is
to mimic a constant external magnetic field.
This could have a significant effect in a well-shielded optical condensate in which the
Lorentz-violating terms may provide a source for symmetry breaking in the system.
Note that the total energy correction due to Lorentz violation grows linearly
with the total number of particles in the condensate
indicating that improved bounds should come with larger condensates.
In order to observe an effect, the effects of Lorentz breaking terms will have to be clearly 
distinguished from an actual stray magnetic field.
This should be possible as the Lorentz-violating terms have constant direction through time
as the apparatus is rotated, while stray magnetic fields will tend to rotate
with the apparatus.
Even a minuscule field may provide the necessary symmetry breaking for an observable
effect.
As an estimate of the experimental sensitivity, the torque applied to the condensate
cloud by the Lorentz-violating background fields is compared to the inertia at some
reasonable angular acceleration that should be observable.
The spin expectation value is taken orthogonal to 
$\vec B \equiv {5 \over 6} \vec B^{(p)} - {1 \over 2}\vec B^{(e)}$ for the maximum effect.
This estimate yields an estimated bound of $|\vec B| \sim m R^2 \alpha$
where $m$ is the mass of a single atom in the condensate, $R$ is the radius of the
condensate, and $\alpha$ is the angular acceleration of the cloud.
Taking $R \sim a_{ho} \sim 1 \mu m$ for a typical condensate with $m \sim 100 GeV$
and $\alpha \sim 10^{-2} s^{-2} = 10^{-19} m^{-2}$ yields an estimated sensitivity
at the $|\vec B| \sim 10^{-29} GeV$ level, comparable\cite{btilde} to the best CPT and Lorentz tests
to date.
Interactions will typically increase the radius of the cloud by one or two orders of 
magnitude\cite{dalfovo} with a corresponding decrease in sensitivity of two to four
orders of magnitude.

\section{Conclusion}

Many of the thermodynamic properties of low-temperature boson gases
remain unaffected by Lorentz violation.
However, when a condensate is present, the specific form of the ground state wave function becomes
important and small symmetry breaking terms can induce collective effects on the 
shape of ground state.
In particular, the main effect on a noninteracting gas is to perturb the ellipsoid slightly.
Given current estimates of the experimental resolution of the wave function shape 
at approximately
the 1\% level, such a direct test is unlikely to yield useful bounds on Lorentz-violation
parameters.
When interactions are dominant, as is the case for most physical condensates, the
kinetic energy terms can often be neglected and the chemical potential absorbs the effects
of the Lorentz-breaking terms leaving the density distribution the same as in the 
conventional case.
The more interesting effects occur when the condensate has multiple
spin components.
The spin couplings act as an effective constant external magnetic field that
can act on the condensate.
These terms have negligible effects on polar condensates for which the expectation
of the spin vanishes.
On the other hand, ferromagnetic condensates couple collectively to the Lorentz
breaking field making it particularly sensitive to the spin couplings.


\begin{thebibliography}{MM1}
\bibitem{fn2}
Throughout the paper,
Lorentz violating terms in the photon sector that could modify the potential 
directly are assumed negligible.  This is a reasonable assumption due to the 
stringent constraints on photon Lorentz violation from astrophysical tests.
\bibitem{btilde}
Note that $B^{(p)}$ and $B^{(e)}$ are denoted elsewhere in
the literature by similar parameters $\tilde b^{(p)}$ and
$\tilde b^{(e)}$.
\end{thebibliography}
\end{document}